# Origin of Non-cubic Scaling Law in Disordered Granular Packing

Chengjie Xia[1], Jindong Li[1], Bingquan Kou[1], Yixin Cao[1], Zhifeng Li[1], Xianghui Xiao[2], Yanan Fu[3], Tiqiao Xiao[3], Liang Hong[1,6], Jie Zhang[1,6], Walter Kob[4], and Yujie Wang[1,5]*

[1]*Department of Physics and Astronomy, Shanghai Jiao Tong University, 800 Dong Chuan Road, Shanghai 200240, China*
[2]*Advanced Photon Source, Argonne National laboratory, 9700 South Cass Avenue, Illinois 60439, USA*
[3]*Shanghai Institute of Applied Physics, Chinese Academy of Sciences, Shanghai 201800, China*
[4]*Laboratoire Charles Coulomb, UMR 5521, University of Montpellier and CNRS, 34095 Montpellier, France*
[5]*Materials Genome Initiative Center, Shanghai Jiao Tong University, 800 Dong Chuan Road, Shanghai 200240, China*
[6] *Institute of Natural Sciences, Shanghai Jiao Tong University, Shanghai 200240, China*

Recent diffraction experiments on metallic glasses have unveiled an unexpected non-cubic scaling law between density and average interatomic distance, which lead to the speculations on the presence of fractal glass order. Using X-ray tomography we identify here a similar non-cubic scaling law in disordered granular packing of spherical particles. We find that the scaling law is directly related to the contact neighbors within first nearest neighbor shell, and therefore is closely connected to the phenomenon of jamming. The seemingly universal scaling exponent around 2.5 arises due to the isostatic condition with contact number around 6, and we argue that the exponent should not be universal.

The origin of dynamic arrest and mechanical rigidity in amorphous materials remains one of the important unresolved questions in condensed matter physics [1-3]. Whether it has a structural origin or is just a dynamic phenomenon remains controversial [4,5]. For metallic glasses, it has long been speculated that dense local packing structures of short-range order serve as the building blocks in these systems [6]. However, how these local structures can be extended to medium or large scales remains at present a mystery due to the existence of geometric frustration or intrinsic chemical disorder [7-9]. Recently, it has been proposed that metallic glasses possess a medium range fractal order, which could rationalize the commonly observed non-cubic scaling law between the position of the first diffraction peak and the bulk density found in neutron and X-ray scattering experiments on these systems [10-13]. The first diffraction peak position is usually associated with the largest inter-plane distance in crystals or the typical nearest neighbor distance in liquids [14-16], and it shows in these systems a power law of exponent 3 as a function of the bulk density since they are three-dimensional by nature. It is therefore surprising that the scaling exponent obtained for metallic glasses under density change induced by either pressure or composition tuning has instead values that are between 2.3 and 2.5 [10-13]. The origin of this anomalous scaling law has been attributed to the presence of a regular or statistical fractal network formed by glass order [6,10,12]. In this picture, the atoms move affinely relative to each other under deformation, and their coherent scattering intensity yields the non-cubic law. However, real metallic glasses are in fact quite compact, while a large-scale fractal structure has zero mass density. Therefore, in order for this picture to be valid, one requires that a substantial amount of atoms exist within the fractal interstitials which do not contribute coherently to the sharp scattering peaks [10]. Another possibility is that the fractal structure only exists up to a finite length scale, above which the system is still homogeneous and three-dimensional [6,12]. These explanations are appealing since they naturally refer to a fractal medium-range glass order, such as percolating icosahedral structures, for metallic glasses, and therefore explain how glass order extends in space. However, the interpretation of the existence of the non-cubic law based on the fractal picture is not without controversy [17], and sometimes one also find deviations from the non-cubic law [18,19].

In this work, we provide microscopic insight to this problem by studying the three-dimensional packing of spherical granular particles, which is a prototypical hard-sphere glass former and has long been considered as a structural model for metallic glasses [20-22]. We identify a non-cubic scaling law in our system, and provide evidences that its origin is local, i.e., without resorting to any fractal structures. Instead, it results from a complex structural evolution of the first-shell neighbors when the packing fraction varies, controlled mainly by the contact neighbors as required by mechanical stability, and the global behavior is a simple statistical average of the local ones. Therefore, such phenomenon is directly related to jamming phenomenon and might be universal near the jamming criticality [23,24]. In the experiment, we used synchrotron X-ray CT techniques to obtain the packing structures of packing with a wide range of packing fractions $\Phi$ [25-34] (see Supplemental Material [35]). In the following, we use the average particle diameter as a unit of length.

The investigation of the non-cubic law can be carried out in both reciprocal and real space. First, we followed the previous scattering experiments on metallic glasses, and studied the structural factor of the packing to investigate the evolution of the peak positions versus $\Phi$. The structure factor is calculated according to $S(q) = \frac{1}{N}\left|\sum_j e^{-i q \cdot r_j}\right|^2$, where $N$ is the number of particles in the probed volume, and shown in Fig. 1(a). The position $q_i$ of $i$th peak is obtained by fitting the peak to a Gaussian function. Previous studies on metallic glasses report scaling behaviors of $\Phi \propto q_i^{D_q(i)}$, with a scaling exponent $D_q$ varies between 3 and 2.5 for the first and second peaks [12]. In our system we find however that, $q_2$ does not change in the whole investigated $\Phi$ range, which corresponding to a very large $D_q(2)$, and the analysis on $q_1$ yields a $D_q(1)$ which is clearly larger than 3 [Fig. 1(b, c)]. Since the interpretation of $S(q)$ is not completely trivial, we have also calculated the pair correlation function $g(r)$ [Fig. 1(d)]. Similar to $S(q)$, we obtain the peak position $p_i$ of the $i$th peak of $g(r)$ by a Gaussian function fit, and determine the scaling behavior $\Phi \propto p_i^{-D_p(i)}$. Note that $p_1 = 1$ for all values of $\Phi$ since the distances between contact neighbors are always 1.0 which yields $D_p(1) = \infty$. $D_p(i)$ decreases from about 5.2 for the second peak to 3.1 for the fourth peak, indicating a cross-over from an anomalous scaling (weak $\Phi$-dependence) at short distances to a normal $\Phi$-dependence on larger length scales [Fig. 1(e-g)]. At first sight the rather different behaviors in both reciprocal and real spaces with respect to the findings in metallic glass systems look surprising. However, we notice that extracting the dimension of a fractal structure from the position of the peak of either $S(q)$ or $g(r)$ is difficult because of the ambiguous and incomplete information they carry [36,37] (see Supplemental Material [35]).

To avoid such ambiguity and understand the essence of non-cubic law on the level of the particles, we develop a more suitable method to define the length scale associated with a fixed number of particle, and then to determine its scaling behavior with $\Phi$. For this, we first sort for each particle its distances to all of its neighbors in ascending order, with the $n$th nearest distance being $r_n$, and then calculate the average neighbor distance of the nearest $n$ neighbors as $R_n = \frac{1}{n}\sum_{i=1}^{n} r_i$. We find that for all $n$ the average distance $\langle R_n \rangle$ follows a scaling relationship, i.e., $\Phi \propto \langle R_n \rangle^{-D_R(n)}$ [see Fig. 2(a) for $n=13$], with an exponent $D_R(n)$ that shows a complex dependence on $n$ i.e., on the length scale considered. In Fig. 2(b) we plot $D_R(n)$ as a function of $\langle r_n \rangle$, where $\langle r_n \rangle$ is the average distance of the $n$th nearest neighbor, which grows for large $n$ like $\langle r_n \rangle \propto n^{1/3}$. Surprisingly we find that $D_R(\langle r_n \rangle)$ shows an oscillatory behavior that is very similar to the one of $g(r)$, and reaches its minimum value $D_R(13) \approx 2.5$ at $\langle r_{13} \rangle \approx 1.37$, which is close to the location of the first valley in $g(r)$ normally considered to be first-shell boundary [Fig. 2(b)]. Thus we see that the scaling exponent of 2.5 found in a series of metallic glasses is reproduced here in our granular system as the minimal value of $D_R(n)$. The figure also shows that for large $n$, $D_R(n)$ converges towards the expected value of 3.0.

The similarity of $D_R(\langle r_n \rangle)$ and $g(r)$ suggests that there exists a close connection between the shell structure of granular packing and the unusual scaling behavior. To elucidate this better, we define $\langle R_{shell,N} \rangle$ as the average distance between the central particle and the particles in the $N$th shell, which are between the $(N-1)$th and the $N$th valleys of $g(r)$, and determine how this distance depends on

$\Phi$: $\Phi \propto \langle R_{shell,N} \rangle^{-D_{shell}(N)}$. Thus $R_{shell,N}$ is a coarse-grained quantity of $r_n$. We find that $D_{shell}(1) \approx 2.5$ and $D_{shell}(N)$ evolves towards 3.0 for large $N$ [left inset of Fig. 2(b)], a result that agrees with previous simulation works on metallic glasses in which the non-cubic scaling law are observed only up to a finite length scale [12].

The oscillatory behavior of $D_R(n)$ shows that neighbors at different distances undergo non-uniform displacements with respect to the central particle when $\Phi$ changes. We thus can single out their contributions to $D_R(n)$ by investigating the behavior of the $n$th nearest neighbors individually, i.e., the scaling relationship of $\Phi \propto \langle r_n \rangle^{-D_r(n)}$ [right inset of Fig 2(b)]. If all particles change their distances to the central particle by the same rate when $\Phi$ varies, $D_r(n)$ should equal to 3 regardless of $n$, while $D_r(n) < (>) 3$ corresponds to an average radial displacement larger (smaller) than a homogeneous one. Within the first shell, we find that $D_r(n) > 3$ for $n \in [1,6]$ and $D_r(n) < 3$ for $n \in [7,13]$, and their overall behavior gives rise to the 2.5 scaling law. This behavior gives us a first hint of how the non-cubic law emerges, which results from the complex non-uniform structural evolution mainly within the first shell as $\Phi$ changes.

To obtain a more specific understanding of the structural origin of the non-cubic law, we determine the $\Phi$-dependence of the local structure within the first shell. For this, we divide the neighbors in the first shell of each particle into two groups. The nearest six ones, i.e., those with $D_r(n) > 3$, and the rest. This classification basically corresponds to the division of quasi-contact and non-contact neighbors owing to the isostatic requirement for mechanical stable granular packings. For each group, we calculate the radial distribution function [Fig. 3(a, b)]. The probability distribution function (PDF) of neighbor-to-center distance $r$ for particles with $n \in [1,6]$ are basically independent of $\Phi$, while the ones for particles with $n \in [7,13]$ show a considerable shift of weight from large to small $r$ as $\Phi$ increases. This observation thus explains why for $n \in [1,6]$ the exponent $D_r(n)$ is large, i.e., no $\Phi$-dependence of $\langle r_n \rangle$, whereas for $n \in [7,13]$ it is small, i.e., strong $\Phi$-dependence of $\langle r_n \rangle$. We also calculate a three point correlation function that gives structural information not accessible from scattering experiments. For this, we measure the angle $\theta$ spanned by the central particle and any two of its neighbors. The distribution of $\theta$ for $n \in [1,6]$ shows a peak at 60° that becomes sharper with increasing $\Phi$ [Fig. 3(c)], which suggests that these particles tend to aggregate to form regular triangles which can further lead to the formation of quasi-regular tetrahedral structures [33,34]. In contrast to this, the distribution of $\theta$ for $n \in [7,13]$ does not show a significant change apart from a slight change in the peak positions [Fig. 3(d)]. The described complex non-affine structural evolution is consistent with the previous observation that the average shape of the Voronoi cells changes from being anisotropic to more isotropic as $\Phi$ increases [33,38,39]. It is this non-affine deformation which induces the deviation from a cubic law between the local packing fraction $\phi$ of the Voronoi cell defined by the first-shell neighbors and their average neighbor-to-center distance $R_{13}$. (We define $\phi$ as the ratio between the volumes of each particle and its Voronoi cell.) Together with the fact that the average $\phi$ is very close to the global $\Phi$, the non-cubic law between $\Phi$ and $\langle R_{13} \rangle$ naturally emerges. Thus above structural analysis supports the local explanation of the non-cubic law irrespective of structural information at medium or long-ranges.

To further justify this local explanation, we make a scatter plot of $\phi$ v.s. $R_{13}$, and fit the scatter plot using $\phi \propto R_{13}^{-d}$ to capture the average behavior (Fig. 4). The scaling exponent can essentially be evaluated by $d = \sigma[\log(\phi)] / \sigma[\log(R_{13})]$, where $\sigma(\cdot)$ represents the standard deviation. Interestingly, $d$ shows an increasing trend from about 2.6 to 2.9 with decreasing $\Phi$ (inset of Fig. 4), which indicates that a local version of the same non-cubic law holds, suggesting that a low-$\Phi$ packing with more liquid-like structure, i.e., smaller contact numbers, has an exponent $d$ closer to 3. This subtle trend is hidden if one fits the global quantities $\Phi$ versus $\langle R_{13} \rangle$ to obtain a single $D_R(13)$. Furthermore, as shown in Fig. 4, the relationship between $\Phi$ and $\langle R_{13} \rangle$ is consistent with the overall local trends, suggesting that the global scaling law is simply an average manifestation of the local non-cubic law between $\phi$ and $R_{13}$ with gradually varying $d$ values.

In the following, we demonstrate that the exponent is closely related to the existence of contact neighbors as required by mechanical stability in granular packing [24,40,41], and is a phenomenon connected to jamming, instead of the fractal glass order as we set out to relate in the first place [34]. This finding is not totally surprising as we recall that even in the work which tried to relate the non-cubic exponent to a presumed fractal glass order in metallic glasses, the anomalous scaling is observed only far below the glass transition temperature, and the potential relationship to jamming is alluded [12].

To illustrate this point, we investigate the dependency of the non-cubic exponents on contact number. Two particles are considered to be in quasi-contact with each other if their surfaces are closer than a cut-off distance around 0.01 of the particle diameter [27,31]. We use *quasi-contact* to identify very close neighbors, which are not necessarily in actual geometric or mechanical contact. In Fig. 5(a), we group the particles based on their local quasi-contact number $z$. The conditional probability distribution of both $\phi$ and $R_{13}$ shift for increasing $z$ values. In each group of particles with fixed $z$, the correlation between $\phi$ and $R_{13}$ can be described by $\phi \propto R_{13}^{-d_z}$, and $d_z$ is again evaluated as the ratio between the standard variances of $\log(\phi)$ and $\log(R_{13})$ for particles with given $z$. As expected, $d_z$ depends on $z$, and increases towards 3 for decreasing $z$. Furthermore, it's intriguing to notice that the relationships between $d_z$ and $z$ are almost identical for all packing with different $\Phi$ [Fig. 5(b)], further confirming that it is a local property. The universal behavior of $d_z$ can therefore describe the $\Phi$-dependence of $d$, even if $d_z$ is a bit smaller than $d$. As we show in the Supplemental Material, this difference originates from the complex inter-dependency between $\phi$, $R_{13}$ and $z$.

In conclusion, we give a local explanation for the origin of a non-cubic law in granular hard-sphere systems, and find it to be related to the phenomenon of jamming instead of a fractal glass structure. Although we do observe in our system the non-cubic scaling laws, the exponents we extract for the peak positions in $S(q)$ and $g(r)$ do not match the ones found in metallic glasses. Thus, our work makes it clear that the non-cubic law might not be universal for both granular and metallic glass systems. For granular systems, the non-universal behavior is presumably due to the presence of friction, which moves system away from the isostatic jamming point. For metallic glasses, since there must be other important parameters (stiffness of potential, covalent bonding, *etc.*), which go beyond the hard-sphere picture and thus will influence this exponent [22]. Also, the rather high temperature at which the scaling law is normally probed in metallic glasses could also influence the exponent. Nevertheless, we believe that a very similar physical mechanism is at work for both systems, since to the first approximation metallic glasses can be described as hard-sphere systems. It is possible in the limit of the isostatic jamming point (with contact number of 6), a universal scaling law of 2.5 indeed exists. This brings us the attention to recent advances in the theory of hard-sphere glasses of a new type of glass transition, the Gardner transition [24,42,43]. This transition happens by breaking the glass metabasins into subbasins by forming a marginal glass. The length scale of this transition is close to that investigated in the current work. It is therefore possible that the scaling exponent identified here is a new structural property of the marginal glass phase or jamming transition, in addition to the cage order parameter or vibration motions normally studied [44]. It is therefore interesting to probe this connection in the future.

The work is supported by the National Natural Science Foundation of China (No. 11175121, 11675110 and U1432111), Specialized Research Fund for the Doctoral Program of Higher Education of China (Grant No. 20110073120073). Experiments were carried out at BL13W1 beamline of the Shanghai Synchrotron Radiation Facility and 2BM beamline of the Advanced Photon Source at Argonne National Laboratory. The use of the Advanced Photon Source is supported by the US Department of Energy, Office of Science, Office of Basic Energy Sciences, under Contract No. DE-AC02-06CH11357. Part of this work was supported by ANR-COMET.

\* Corresponding author.

yujiewang@sjtu.edu.cn

[1] P. G. Debenedetti and F. H. Stillinger, Nature (London) **410**, 259 (2001).
[2] L. Berthier and G. Biroli, Rev. Mod. Phys. **83**, 587 (2011).
[3] K. Binder and W. Kob, *Glassy materials and disordered solids: An introduction to their statistical mechanics* (World Scientific, 2011).
[4] D. Chandler and J. P. Garrahan, Annu. Rev. Phys. Chem. **61**, 191 (2010).
[5] S. Karmakar, C. Dasgupta, and S. Sastry, Annu. Rev. Condens. Matter Phys. **5**, 255 (2014).
[6] Y. Q. Cheng and E. Ma, Prog. Mater. Sci. **56**, 379 (2011).
[7] D. B. Miracle, Nat. Mater. **3**, 697 (2004).
[8] H. W. Sheng, W. K. Luo, F. M. Alamgir, J. M. Bai, and E. Ma, Nature (London) **439**, 419 (2006).
[9] Y. C. Hu, F. X. Li, M. Z. Li, H. Y. Bai, and W. H. Wang, Nat. Commun. **6**, 8310 (2015).
[10] D. Ma, A. D. Stoica, and X.-L. Wang, Nat. Mater. **8**, 30 (2009).
[11] Q. Zeng *et al.*, Phys. Rev. Lett. **112**, 185502 (2014).
[12] D. Z. Chen, C. Y. Shi, Q. An, Q. Zeng, W. L. Mao, W. A. Goddard, and J. R. Greer, Science **349**, 1306 (2015).
[13] Q. Zeng *et al.*, Proc. Natl. Acad. Sci. U. S. A. **113**, 1714 (2016).
[14] H. F. Poulsen, J. A. Wert, J. Neuefeind, V. Honkimaki, and M. Daymond, Nat. Mater. **4**, 33 (2005).
[15] A. R. Yavari *et al.*, Acta Mater. **53**, 1611 (2005).
[16] T. C. Hufnagel, R. T. Ott, and J. Almer, Phys. Rev. B **73**, 064204 (2006).
[17] P. Chirawatkul, A. Zeidler, P. S. Salmon, S. Takeda, Y. Kawakita, T. Usuki, and H. E. Fischer, Phys. Rev. B **83**, 014203 (2011).
[18] O. F. Yagafarov, Y. Katayama, V. V. Brazhkin, A. G. Lyapin, and H. Saitoh, Phys. Rev. B **86**, 174103 (2012).
[19] A. Zeidler and P. S. Salmon, Phys. Rev. B **93**, 214204 (2016).
[20] F. C. Frank, Proc. R. Soc. (London) **215**, 43 (1952).
[21] J. Bernal and J. Mason, Nature (London) **188**, 910 (1960).
[22] K. Zhang, M. Fan, Y. H. Liu, J. Schroers, M. D. Shattuck, and C. S. O'Hern, J. Chem. Phys. **143**, 184502 (2015).
[23] C. S. O'Hern, L. E. Silbert, A. J. Liu, and S. R. Nagel, Phys. Rev. E **68**, 011306 (2003).
[24] P. Charbonneau, J. Kurchan, G. Parisi, P. Urbani, and F. Zamponi, Nat. Commun. **5**, 3725 (2014).
[25] G. T. Seidler, G. Martinez, L. H. Seeley, K. H. Kim, E. A. Behne, S. Zaranek, B. D. Chapman, S. M. Heald, and D. L. Brewe, Phys. Rev. E **62**, 8175 (2000).
[26] P. Richard, P. Philippe, F. Barbe, S. Bourles, X. Thibault, and D. Bideau, Phys. Rev. E **68**, 020301(R) (2003).
[27] T. Aste, M. Saadatfar, and T. Senden, Phys. Rev. E **71**, 061302 (2005).
[28] M. Hanifpour, N. Francois, S. M. V. Allaei, T. Senden, and M. Saadatfar, Phys. Rev. Lett. **113**, 148001 (2014).
[29] Y. Fu, Y. Xi, Y. Cao, and Y. Wang, Phys. Rev. E **85**, 051311 (2012).
[30] Y. Cao, B. Chakrabortty, G. C. Barker, A. Mehta, and Y. Wang, Europhys. Lett. **102**, 24004 (2013).
[31] C. Xia, K. Zhu, Y. Cao, H. Sun, B. Kou, and Y. Wang, Soft Matter **10**, 990 (2014).
[32] J. Li, Y. Cao, C. Xia, B. Kou, X. Xiao, K. Fezzaa, and Y. Wang, Nat. Commun. **5**, 5014 (2014).
[33] C. Xia, Y. Cao, B. Kou, J. Li, Y. Wang, X. Xiao, and K. Fezzaa, Phys. Rev. E **90**, 062201 (2014).
[34] C. Xia, J. Li, Y. Cao, B. Kou, X. Xiao, K. Fezzaa, T. Xiao, and Y. Wang, Nat. Commun. **6**, 8409 (2015).
[35] See Supplemental Material at [] for details regarding.
[36] N. Mattern, M. Stoica, G. Vaughan, and J. Eckert, Acta Mater. **60**, 517 (2012).
[37] L. Hongbo, W. Xiaodong, C. Qingping, Z. Dongxian, Z. Jing, H. Tiandou, M. Ho-kwang, and J. Jian-Zhong, Proc. Natl. Acad. Sci. U. S. A. **110**, 10068 (2013).
[38] P. Richard, J. P. Troadec, L. Oger, and A. Gervois, Phys. Rev. E **63**, 062401 (2001).
[39] G. E. Schröder-Turk, W. Mickel, M. Schröter, G. W. Delaney, M. Saadatfar, T. J. Senden, K. Mecke, and T. Aste, Europhys. Lett. **90**, 34001 (2010).
[40] A. J. Liu and S. R. Nagel, Nature (London) **396**, 21 (1998).
[41] C. Song, P. Wang, and H. A. Makse, Nature (London) **453**, 629 (2008).
[42] G. Biroli and P. Urbani, Nat. Phys. **12**, 1130 (2016).
[43] Y. Jin and H. Yoshino, arXiv:1610.07301v2.
[44] L. Berthier, P. Charbonneau, Y. L. Jin, G. Parisi, B. Seoane, and F. Zamponi, Proc. Natl. Acad. Sci. U. S. A. **113**, 8397 (2016).

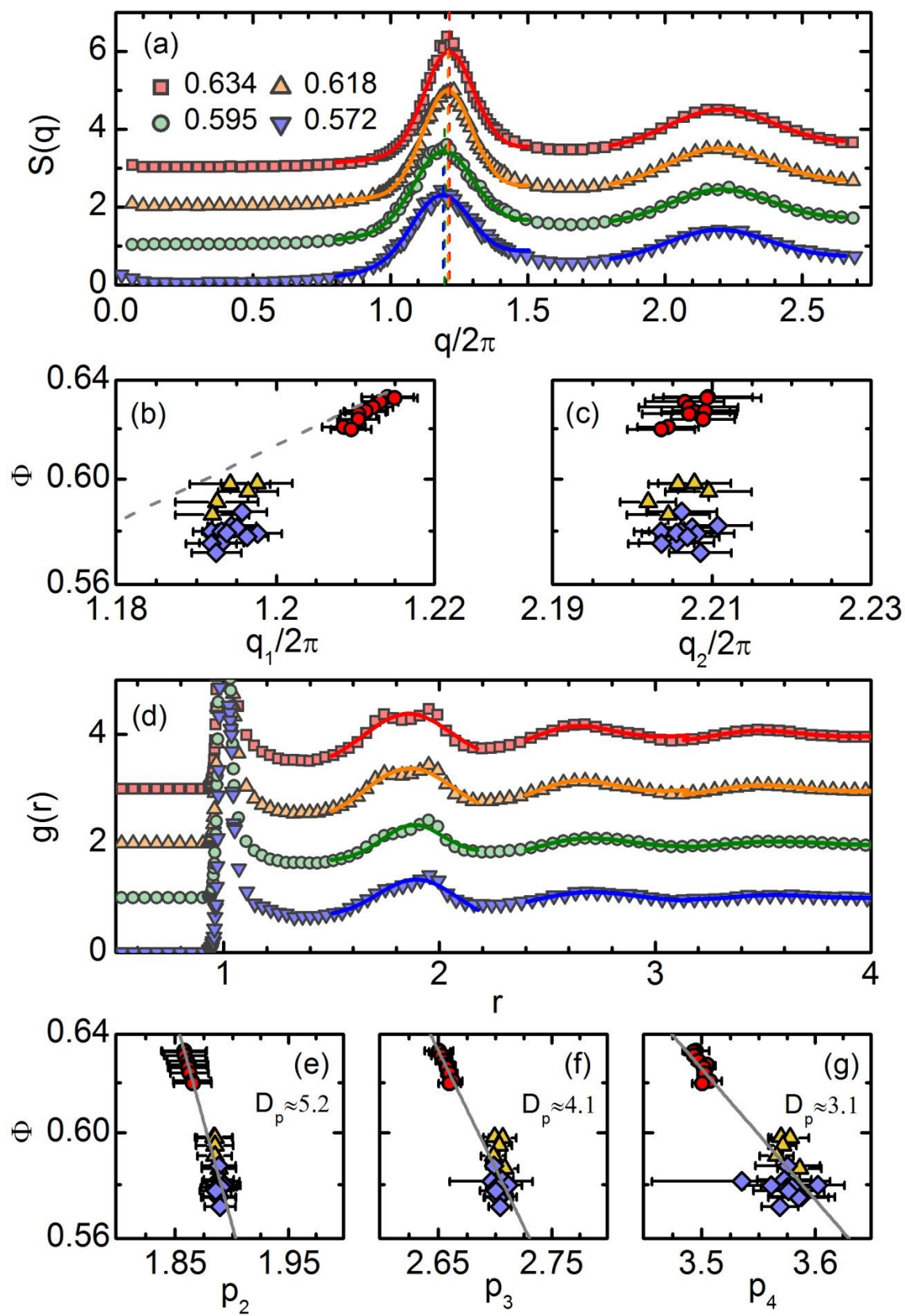

FIG. 1. (a) Structure factors and (d) pair correlation functions for four different packing fractions $\Phi$ (see legend). The data are shifted vertically for clarity. Solid lines are Gaussian fits to the peaks. Dashed lines mark the peak positions. (b, c) Peak positions of the first and second peak of $S(q)$ for all packing. The size of the error-bar of $\Phi$ is smaller than the symbol size. The dashed line in (b) represents $\Phi \propto q_1^3$. (e-g) Peak positions of the second to fourth peak of $g(r)$. The solid lines is the fit of the form $\Phi \propto p_i^{-D_p(i)}$, with $D_p(2) = 5.2 \pm 0.5$, $D_p(3) = 4.1 \pm 0.5$ and $D_p(4) = 3.1 \pm 0.3$. Different symbols in (b, c, e-g) represents three different packing preparation protocols as explained in Fig. 2(a).

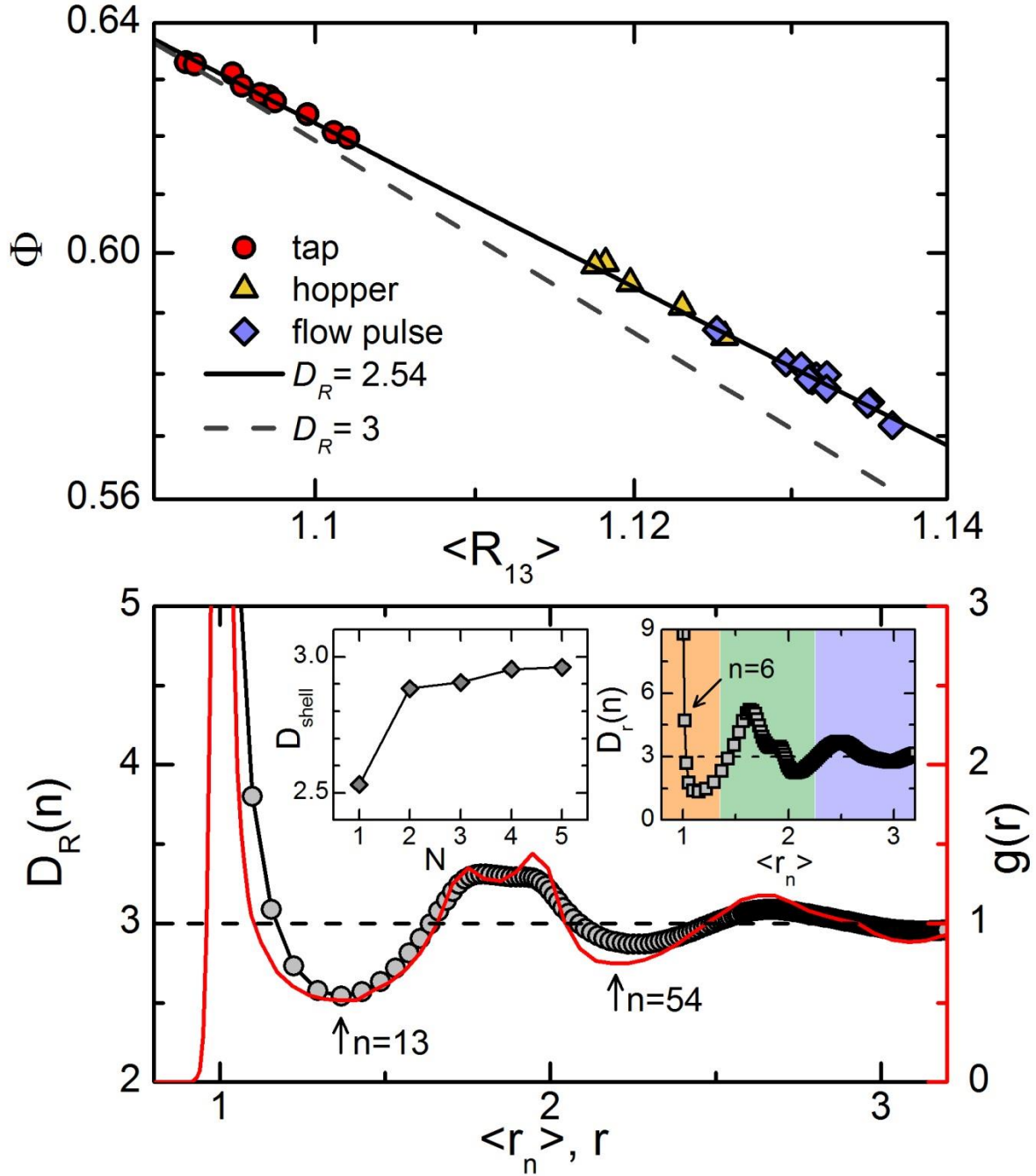

FIG. 2. (a) $\Phi$ versus $\langle R_{13} \rangle$. Different symbols represent different packing preparation protocols: Tapping (circles), hopper (triangles) and flow pulse (diamonds). A clear non-cubic law can be identified through the fit $\Phi \propto \langle R_{13} \rangle^{-D_R(13)}$ with $D_R(13) = 2.54 \pm 0.03$ (solid line) with a cubic law (dashed line) for comparison. (b) $D_R(n)$ versus $\langle r_n \rangle$ (symbols, left axis) and $g(r)$ (line, right axis) for packing with

$\Phi = 0.634$. The dashed line represents $D_R = 3$. $n = 13$ and $n = 54$ are marked as the boundaries of the first two shells. Left inset: $D_{shell}(N)$ versus shell number $N$. Right inset: $D_r(n)$ versus $\langle r_n \rangle$. The dashed line represents $D_r = 3$, and the background colors separate the different shells.

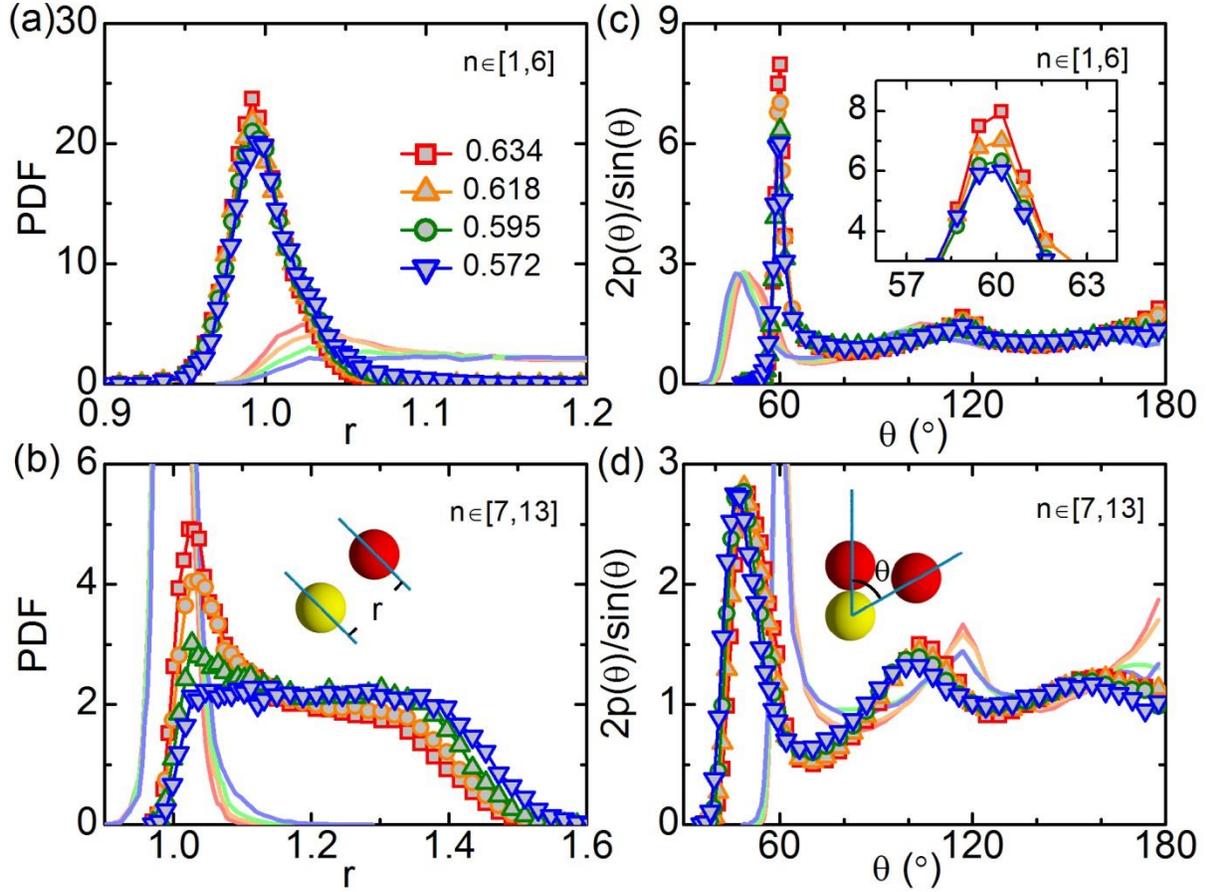

FIG.3. PDFs of neighbor-to-center distances $r$ (a, b) and angle $\theta$ (c, d) for particles with $n \in [1,6]$ (a, c) and $n \in [7,13]$ (b, d) for four different packing fractions given in the legend. The inset in (c) is a zoom onto the peak at $60°$. In (c, d), $\sin(\theta)/2$ is a normalization factor. For the sake of comparison, we plot in each panel the PDF for the other group of particles (solid lines). Panels (b) and (d) also show a schematic picture of the definitions of $r$ and $\theta$.

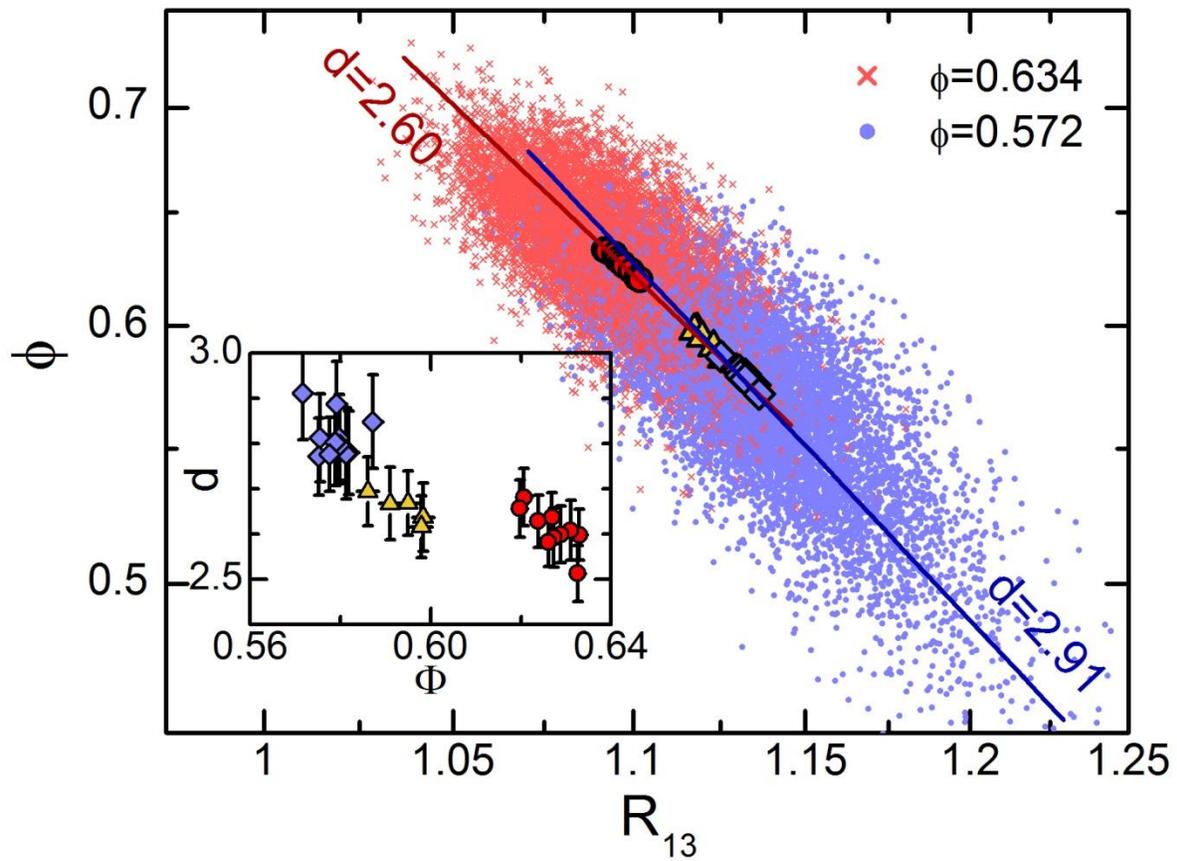

FIG.4. Scatter plot of $\phi$ and $R_{13}$ for our densest (cross) and loosest packing (dot). The solid lines represent $\phi \propto R_{13}^{-d}$, with $d = 2.60 \pm 0.02$ for $\Phi = 0.634$ and $d = 2.91 \pm 0.03$ for $\Phi = 0.572$. The global average values of $\Phi$ and $\langle R_{13} \rangle$ are also shown. Inset: $d$ as a function of $\Phi$.

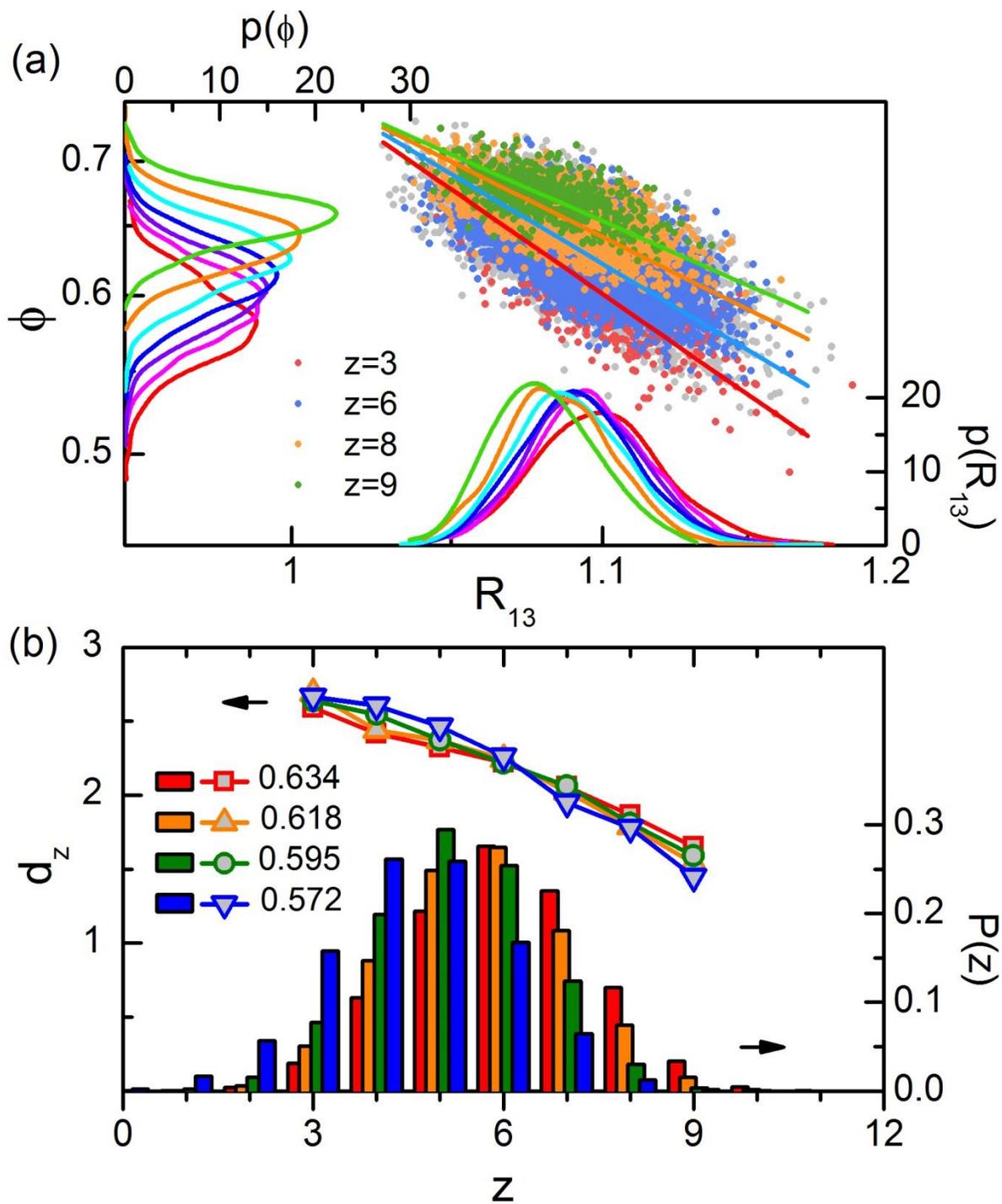

FIG.5. (a) Scatter plot of $\phi$ and $R_{13}$ for particles of packing with $\Phi = 0.634$ with different $z$. For clarity, only particles with $z = 3, 6, 8, 9$ are colored and the rest ones are plotted in gray dots. The solid

lines represent $\phi \propto R_{13}^{-d_z}$. Conditional PDFs of $\phi$ (top axis) and $R_{13}$ (right axis) for different $z$ are also shown. (b) $d_z$ as a function of $z$ for four different packing (left axis), and the probability distribution of $z$ for the same four packing (right axis).

# Supplemental Material for

# Origin of Non-cubic Scaling Law in Disordered Granular Packing

Chengjie Xia[1], Jindong Li[1], Bingquan Kou[1], Yixin Cao[1], Zhifeng Li[1], Xianghui Xiao[2], Yanan Fu[3], Tiqiao Xiao[3], Liang Hong[1,6], Jie Zhang[1,6], Walter Kob[4], and Yujie Wang[1,5]*

[1]*Department of Physics and Astronomy, Shanghai Jiao Tong University, 800 Dong Chuan Road, Shanghai 200240, China*
[2]*Advanced Photon Source, Argonne National laboratory, 9700 South Cass Avenue, Illinois 60439, USA*
[3]*Shanghai Institute of Applied Physics, Chinese Academy of Sciences, Shanghai 201800, China*
[4]*Laboratoire Charles Coulomb, UMR 5521, University of Montpellier and CNRS, 34095 Montpellier, France*
[5]*Materials Genome Initiative Center, Shanghai Jiao Tong University, 800 Dong Chuan Road, Shanghai 200240, China*
[6]*Institute of Natural Sciences, Shanghai Jiao Tong University, Shanghai 200240, China*

## A. EXPERIMENTAL DETAILS

The granular packing is composed of glass beads (Duke Scientific, USA) with $200\,\mu m$ particle diameters and around 3% polydispersity. The packing are prepared using three different methods, i.e., tapping, hopper deposition, and flow pulse, to obtain a broad range of global packing fractions $\Phi$ from 0.57 to 0.64 [34]. In tapping and flow pulse protocols, $\Phi$ is tuned by changing the tapping intensity or the flow velocity respectively, while in the hopper deposition protocol, particles gradually drain from a slowly lifting hopper and form a pile with $\Phi \approx 0.59$. In both tapping and flow pulse protocols, samples are prepared to reach the steady state after long sequences of external excitations. The packing structures are obtained using synchrotron X-ray CT techniques, with a spatial resolution of $5.5\,\mu m$. The centroids and sizes of all particles are obtained through a series of well-established image processing techniques, in which the centroid of each particle is calculated as the average position of all its voxels in the tomographic image, with a precision about 0.3% average particle diameter. In this work, only particles three-diameters away from the container boundary are used in the analysis, which leaves approximately 17000 particles for each packing.

## B. PARTICLE NUMBER ASSOCIATED WITH $g(r)$

The derivation of a fractal spatial mass distribution relies on relating the change of global volume (by either chemical substitution or compression) and the change of linear dimension for clusters with *fixed particle number (mass)*. This is clearly violated when we work only with the peaks of $S(q)$ or $g(r)$ since the quasi-contact number $z$ for the first peak of $g(r)$ changes from about 4.5 to 6 in our $\Phi$ range, and the number of particles lies within the first valley (the coordination number $N$) also increases from about 11.5 to 12.5 (Fig. S1).

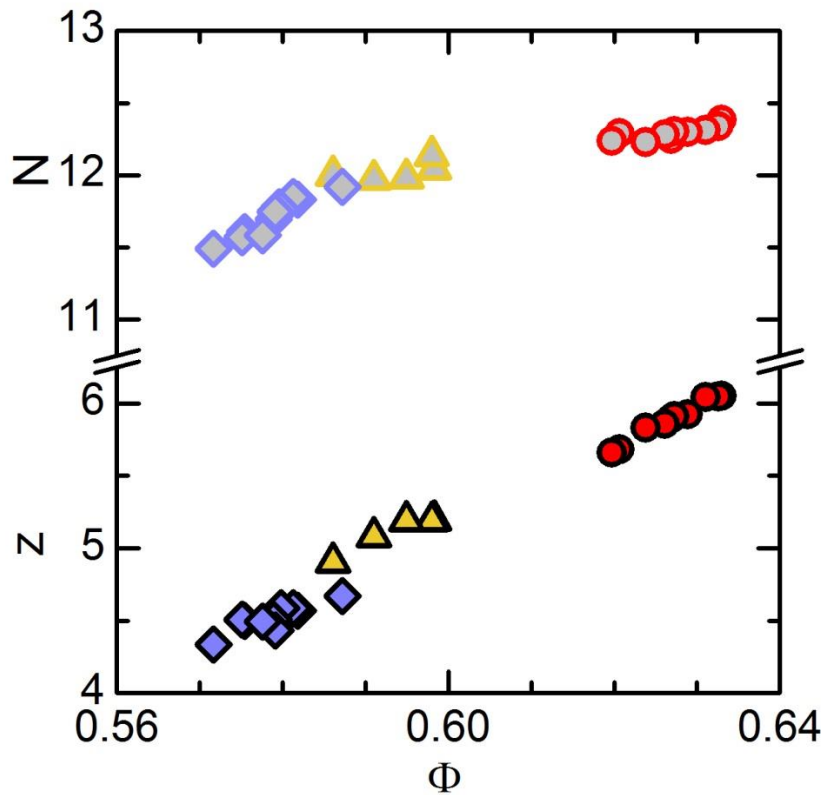

FIG. S1. Average quasi-contact number $z$ and coordination number $N$ for packing with different $\Phi$.

## C. RELATIONSHIP BETWEEN $d$ AND $d_z$

In the following we give a rough explanation why $d$ and $d_z$ differ, and elucidate how particles with different $z$ give rise to the overall relationship between $\phi$ and $R_{13}$ when one takes the average over $z$. In a phenomenological way, we express the local packing fraction $\phi = \phi(z, R_{13}, ...)$ as a function of both $z$ and $R_{13}$, in addition to the rest of possible relevant control parameters. Note that here $z$ and $R_{13}$ are not really independent variables, since these quantities are correlated. We also mention that a simple cubic law can be expected if the contact number is irrelevant, such as it is the case in systems like simple liquids. The scaling exponent of the overall relationship $\phi \propto R_{13}^{-d}$ is quantified by $d = -\left\langle \dfrac{d \log(\phi)}{d \log(R_{13})} \right\rangle_z$ where $\langle \cdot \rangle_z$ stands for a statistical average with different $z$ values. Since $R_{13}$ depends on $z$, we can write

$$\frac{d \log(\phi)}{d \log(R_{13})} = \left.\frac{\partial \log(\phi)}{\partial \log(R_{13})}\right|_z + \frac{\partial \log(\phi)}{\partial z} \frac{\partial z}{\partial \log(R_{13})} + ...,$$

and thus we have $d = \langle d_z \rangle + \langle d_2 \rangle + d_{rem}$, where $\langle d_z \rangle = \sum_z P(z) d_z$, $d_z = -\left.\dfrac{\partial \log(\phi)}{\partial \log(R_{13})}\right|_z$ by definition, $\langle d_2 \rangle = -\left\langle \dfrac{\partial \log(\phi)}{\partial z} \dfrac{\partial z}{\partial \log(R_{13})} \right\rangle_z$, and $d_{rem}$ denotes all remaining contribution from the rest control parameters. $\langle d_z \rangle$ increases from 2.2 to 2.5 for decreasing $\Phi$ as a result of decreasing $z$ (Fig. S2). In Fig. S3(a), we show the distribution of $\phi$ for different values of $z$ for packing with $\Phi = 0.634$, and in Fig. S3(b) the distribution of $R_{13}$ for different $z$. For clarity the curves for different $z$ have been shifted vertically. The fact that the position of their peaks shifts linearly with $z$ on logarithmic scale (see straight lines) shows that these distributions depend linearly on $z$, i.e., we can write $\log(\phi) = b_z + k_z z$ and

$z = b_R + k_R \log(R_{13})$. Here $k_z$ and $k_R$ are the slopes. So the product $\langle d_2 \rangle = -\left\langle \dfrac{\partial \log(\phi)}{\partial z} \dfrac{\partial z}{\partial \log(R_{13})} \right\rangle_z$ can be approximated by $\langle d_2 \rangle = -k_z k_R$. It turns out that $\langle d_2 \rangle$ lies in the range between 0.15 to 0.3, and $d_{rem} = d - \langle d_z \rangle - \langle d_2 \rangle \approx 0.1$ for all different packing. Now it's clear that the major contribution to the deviation of exponent from 3 is due to the existence of the contact neighbors, while the contributions from the complex inter-dependency between $\phi$, $z$ and $R_{13}$, and the remaining high-order terms contribute a minor role. Therefore, it is not surprising to observe a seemingly universal 2.5 scaling exponent in so many systems, when our hard-sphere granular system or metallic glasses are close to the isostatic point of the jamming transition with contact number being around 6.

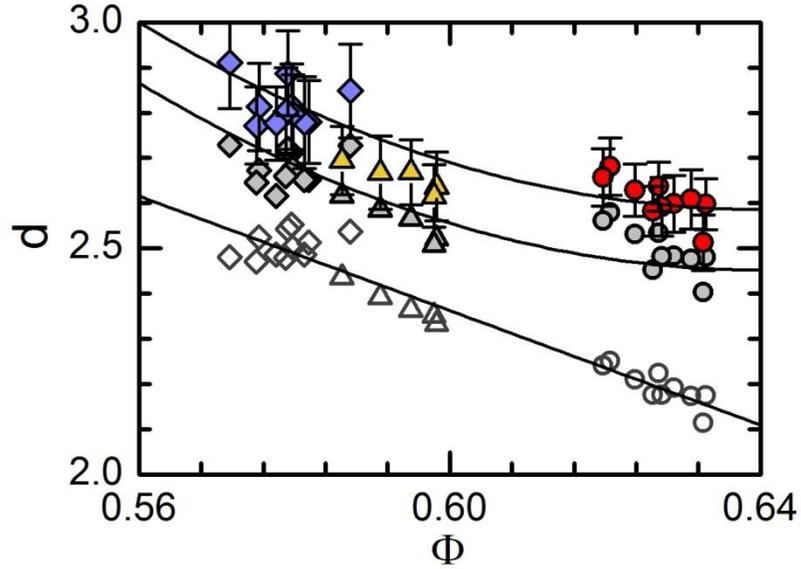

FIG. S2. $d$ (colored symbols), $\langle d_z \rangle$ (open symbols) and $\langle d_z \rangle + \langle d_2 \rangle$ (gray symbols) for packing with different $\Phi$. The solid lines are guides to eye.

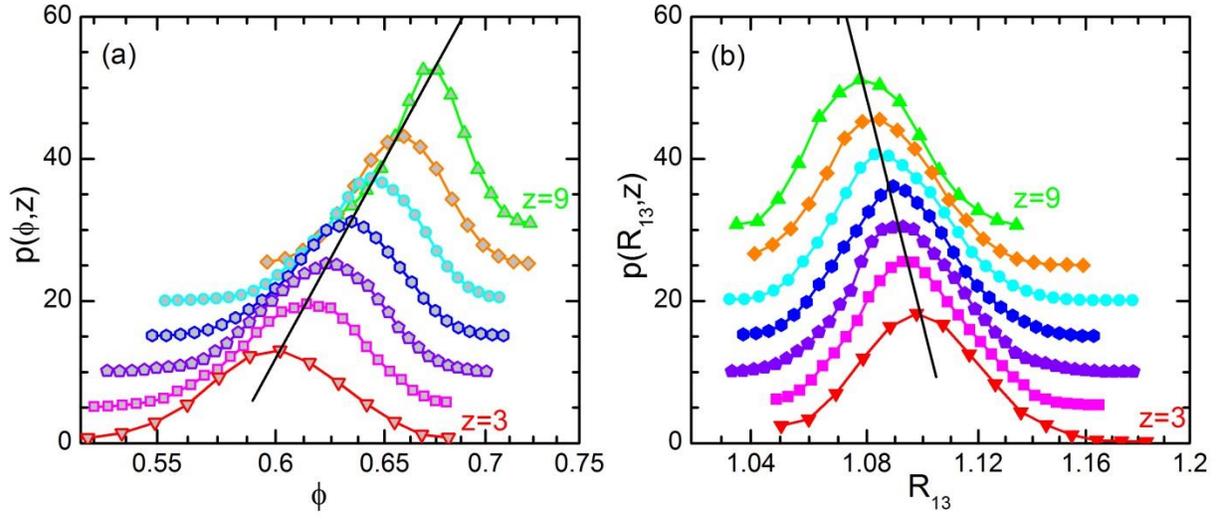

FIG. S3. PDFs of (a) $\phi$ and (b) $R_{13}$ for different values of $z$, for packing with $\Phi = 0.634$. The curves for different $z$ have been shifted vertically. The straight lines show the linear dependency of the distributions on $z$.